\documentclass{ieeeaccess}
\usepackage{cite}
\usepackage{amsmath,amssymb,amsfonts}
\usepackage{algorithmic}
\usepackage{graphicx}
\usepackage{textcomp}
\usepackage{cite} 
\usepackage{bm}

\usepackage[colorlinks=true, linkcolor=blue, citecolor=blue, urlcolor=blue]{hyperref}
\usepackage{amsmath}
\usepackage{booktabs}
\usepackage{graphicx}       
\usepackage{array}          
\usepackage{booktabs}       
\usepackage{multicol}       
\usepackage{multirow}       
\usepackage{amsmath}        
\usepackage{float}          
\usepackage{enumitem}

\usepackage[ruled,vlined,noend]{algorithm2e} 
\usepackage{xcolor}
\usepackage{soulutf8}

\makeatletter
\AtBeginDocument{\DeclareMathVersion{bold}
\SetSymbolFont{operators}{bold}{T1}{times}{b}{n}
\SetSymbolFont{NewLetters}{bold}{T1}{times}{b}{it}
\SetMathAlphabet{\mathrm}{bold}{T1}{times}{b}{n}
\SetMathAlphabet{\mathit}{bold}{T1}{times}{b}{it}
\SetMathAlphabet{\mathbf}{bold}{T1}{times}{b}{n}
\SetMathAlphabet{\mathtt}{bold}{OT1}{pcr}{b}{n}
\SetSymbolFont{symbols}{bold}{OMS}{cmsy}{b}{n}
\renewcommand\boldmath{\@nomath\boldmath\mathversion{bold}}}
\makeatother

\def\BibTeX{{\rm B\kern-.05em{\sc i\kern-.025em b}\kern-.08em
    T\kern-.1667em\lower.7ex\hbox{E}\kern-.125emX}}

\begin{document}
\history{This work has been published in IEEE Access, Volume 13, on 5 March 2025. Published version available at: https://ieeexplore.ieee.org/document/10910185}
\doi{10.1109/ACCESS.2025.3548031}

\title{Efficient Computation of Collatz Sequence Stopping Times: A Novel Algorithmic Approach}
\author{\uppercase{Eyob Solomon Getachew}\authorrefmark{1},
\uppercase{Beakal Gizachew Assefa}\authorrefmark{2}}

\address[1]{School of Information Technology and Engineering, Addis Ababa Institute of Technology, Addis Ababa, Ethiopia (e-mail: eyob.solomon@aait.edu.et)}
\address[2]{School of Information Technology and Engineering, Addis Ababa Institute of Technology, Addis Ababa, Ethiopia (e-mail: beakal.Gizachew@aait.edu.et)}


\corresp{Corresponding author: Eyob Solomon Getachew (e-mail: eyob.solomon@aait.edu.et).}

\begin{abstract}
The Collatz conjecture, which posits that any positive integer will eventually reach 1 through a specific iterative process, is a classic unsolved problem in mathematics. This research focuses on designing an efficient algorithm to compute the stopping time of numbers in the Collatz sequence, achieving significant computational improvements. By leveraging structural patterns in the Collatz tree, the proposed algorithm minimizes redundant operations and optimizes computational steps. Unlike prior methods, it efficiently handles extremely large numbers without requiring advanced techniques such as memoization or parallelization. Experimental evaluations confirm computational efficiency improvements of approximately 28\% over state-of-the-art methods. These findings underscore the algorithm’s scalability and robustness, providing a foundation for future large-scale verification of the conjecture and potential applications in computational mathematics.
\end{abstract}


\titlepgskip=-21pt

\maketitle

\section{Introduction}
\label{sec:introduction}
\PARstart{T}{he} Collatz conjecture, also known as the 3n+1 problem, is a long-standing open question in mathematics that continues to captivate researchers due to its simplicity and the profound computational challenges it poses. Proposed by Lothar Collatz in 1937, the conjecture posits that any positive integer will eventually reach the cycle \(4, 2, 1\) through a specific iterative process: dividing by 2 if the number is even, or applying the transformation \(3n + 1\) if odd \cite{kosova2023collatz}. Despite its simple formulation, no formal proof has been found, though extensive computational testing has verified the conjecture for a wide range of numbers, including values as large as \(2^{100000}-1\) \cite{ren2018collatz}. 

The enduring fascination with the Collatz conjecture stems from its ability to bridge elementary number theory and advanced computational techniques, making it both accessible and challenging. The conjecture's appeal lies not only in its theoretical depth but also in the computational challenges it presents. These challenges have spurred the development of algorithms to analyze the stopping time of numbers in the Collatz sequence—the number of steps required for a given number to reach 1 \cite{venkatesulu2020verification}. A precise understanding of stopping times is critical for large-scale verification efforts and has implications for optimizing computational strategies.

This paper introduces a novel algorithm for calculating the stopping time of numbers in the Collatz sequence. The proposed algorithm achieves significant improvements in computational efficiency by leveraging structural patterns within the Collatz tree to minimize redundant operations. Key optimizations include identifying and exploiting repetitive patterns within the tree, allowing the algorithm to bypass unnecessary computations. Experimental evaluations demonstrate that the algorithm consistently outperforms existing state-of-the-art methods, achieving computational efficiency improvements of approximately 28\% across various input sizes, ranging from small to extremely large numbers. 

By advancing algorithmic techniques for the Collatz conjecture, this research contributes to the broader field of computational mathematics, providing a foundation for further exploration and potential applications of these optimized methods in other domains.

\section{Literature Review}

The literature on the Collatz conjecture spans diverse perspectives and approaches, broadly categorized into three main areas: (A) theoretical insights and structural analysis, (B) computational methods and algorithmic efficiency, and (C) practical applications and implications. Theoretical studies examine the conjecture's mathematical foundations and the structural intricacies of its behavior. These studies aim to uncover patterns, propose generalizations, and assess probabilistic models. Computational advancements explore algorithmic designs that push the limits of numeric verification. Finally, practical applications leverage the conjecture’s iterative properties for real-world problems, such as cryptography, blockchain consensus, and digital watermarking. The following sections provide a detailed review of these categories.

\subsection*{A. Theoretical Insights and Structural Analysis}
Recent investigations into the Collatz conjecture have employed both structural and probabilistic approaches, seeking patterns and behaviors that might help reveal or approximate a proof. Techniques such as bottom-up analysis, sequence generalization, probabilistic modeling, and logical frameworks have been explored to better understand convergence properties, stopping times, and potential counterexamples.

The bottom-up approach constructs Collatz sequences in reverse, starting from \( n = 1 \) and applying inverse transformations to map all integers back to this origin. By doubling even numbers and modifying odd numbers with \( (n-1)/3 \), this method forms a tree-like structure that visualizes convergence paths to 1. This analysis yields algebraic formulas generalizing the sequence's behavior, offering insight into its convergence properties \cite{abascal2024bottom}.

An approximation method analyzes stopping times in the Collatz sequence by defining transformations as \( T(m) = m / 2 \) for even \( m \) and \( T(m) = (3m + 1) / 2 \) for odd \( m \). This method introduces probabilistic bounds, showing that average stopping times fall within a logarithmic framework and providing insights into potential asymptotic behaviors of the conjecture. The model provides lower bounds, contributing insights into the asymptotic behavior of stopping times if the conjecture holds \cite{inselmann2024approximation}.

A framework that conceptualizes the Collatz sequence as a concurrent program with "convergence stairs" allows numbers to be grouped by steps needed to reach 1. Infinite binary trees represent each "stair," which visualizes convergence without exhaustive backward tracing. This structured approach offers a layered view of the conjecture, potentially aiding in analyzing complex non-linear systems \cite{ebnenasir2024specifying}.

The generalization \(3n + 3k\), where \(k \in \mathbb{N}\), extends the Collatz sequence’s behavior. For \(k = 0\), the sequence mirrors the familiar 3n + 1 pattern with the cycle 4, 2, 1. Higher values of \(k\) produce new periodic sequences, with formulas detailing their stopping times, presenting a broader framework for Collatz-like transformations \cite{boulkaboul20223}.

An extensive review of computational and theoretical efforts to find Collatz counterexamples shows no success in uncovering a counterexample. The analysis also suggests that iterative reduction strategies might either contribute to proving the conjecture or demonstrate its potential as an unprovable truth in mathematics, enriching its theoretical study \cite{clay2023long}.

A statistical analysis assesses the probability of infinite stopping times in the Collatz sequence, which would contradict the conjecture. The results indicate that the likelihood of divergence is astronomically low, analogous to selecting a specific atom from the entire observable universe, reinforcing the assumption that all sequences eventually converge to 1 \cite{nicola2024some}.

A logical and probabilistic approach to proving the Collatz conjecture addresses the absence of non-trivial infinite loops and posits that all natural numbers eventually converge to 1. A probabilistic model further supports this by demonstrating that the likelihood of divergence approaches zero as iterations increase, providing structured reasoning for the conjecture’s validity \cite{zhou2024proof}.

\subsection*{B. Computational Methods and Algorithmic Efficiency}
Significant computational advances have allowed the Collatz conjecture to be empirically verified for extremely large numbers, employing a variety of algorithmic approaches to push the limits of computational capabilities. Techniques such as bitwise operations, binary representation, and automata-based processing have emerged as essential tools in this verification.

Leveraging bitwise operations and disk-based storage, a high-capacity algorithm represents numbers in binary to efficiently perform 3x+1 computations on a bit-by-bit basis. This method has enabled verification for numbers up to 100,000 bits, confirming the conjecture for values up to \(2^{100000} - 1\) and demonstrating its effectiveness for unprecedented numeric ranges \cite{ren2018collatz}. Similarly, another binary-based algorithm transforms integers to binary form, applying bitwise operations and reverse shifts to verify the conjecture for sequences up to 3000 bits. This approach confirms bounded growth, with no cycles beyond expected bounds, reinforcing the conjecture’s validity \cite{venkatesulu2020verification}.

In addition to bitwise techniques, a set of novel algorithms and theorems analyze the structural relationships within Collatz sequences. These tools include a directed graph visualization, an iteration analyzer, and a peak value finder that collectively explore the relationship between natural numbers, peak values, and iteration counts. Such tools provide computational insights into the conjecture’s complex structure, which could support eventual proof efforts \cite{schwob2021novel}.

Automata-based verification offers further computational efficiency by processing transformations in batches rather than individually. This method utilizes a “Collatz tree” organized by residue classes to enable direct processing of certain values without full transformations. By segmenting repetitive patterns, this design significantly optimizes the verification process and provides a scalable framework for future work \cite{ren2019fast}.

\subsection*{C. Applications and Practical Implications}
Beyond mathematical verification, the conjecture’s iterative properties have found applications in various areas, including blockchain consensus mechanisms, cryptographic security, digital watermarking, image encryption, and medical diagnostics.

Studies on Proof-of-Collatz Conjecture (PCC) algorithms in blockchain demonstrate the conjecture’s potential as a consensus mechanism, effectively replacing traditional Proof-of-Work (PoW). By leveraging the unpredictable stopping times of Collatz sequences, the PCC approach achieves stable execution times and reduces computational costs, offering an efficient alternative to PoW in private blockchain environments \cite{aljassas2019performance}.

In cryptographic security, Collatz sequences are applied in both audio and image encryption, as well as hash function design. Collatz-based algorithms in audio encryption convert speech signals into unintelligible forms using variable-length encoding, achieving high entropy and low data correlation with the original audio to enhance transmission security \cite{renza2019high}. Similarly, an encrypted audio dataset leverages these sequence transformations to provide high-entropy audio signals that obscure original content, creating a valuable resource for cryptanalysis testing \cite{renza2019encrypted}. For image encryption, a chaotic logistic map-based scheme incorporates Collatz sequences to generate high-entropy encryption keys, significantly improving throughput and resilience against cryptographic attacks \cite{al2024generating}. Additionally, a Collatz-based hash function merges chaotic properties with sequence unpredictability, achieving enhanced randomness, strong diffusion, and improved collision resistance, outperforming traditional hash functions like SHA-2 and SHA-3 \cite{rasool2023collatz}.

In digital watermarking, a novel method embeds watermarks by calculating the Collatz degree of each pixel in an image, enhancing visual quality and providing a secure, low-complexity solution suitable for real-time applications. This approach supports efficient image authentication through a pseudo-random number generator for block-wise embedding \cite{tuncer2022novel}. Another image encryption scheme transforms images into encrypted audio by encoding pixel values through Collatz transformations. This combined scrambling and diffusion process yields high entropy and strong key sensitivity, offering robust protection against cryptographic attacks while allowing reversible recovery \cite{ballesteros2018novel}.

In medical diagnostics, the Collatz pattern is applied to schizophrenia detection by transforming EEG signals into unique feature vectors. Using maximum absolute pooling and iterative neighborhood component analysis (INCA), this technique selects clinically significant features and classifies them with k-nearest neighbors (kNN), providing a high-accuracy, efficient tool for automated EEG-based diagnosis \cite{baygin2021automated}.

\section{Methodology}

This section outlines the methodology for developing and analyzing an efficient algorithm to compute stopping times in the Collatz sequence. It begins with an exploration of structural patterns in the Collatz tree, which serve as the foundation for algorithmic optimization. The proposed algorithm is then described, followed by an analysis of its computational efficiency and correctness. The algorithm’s efficiency is compared with state-of-the-art methods, followed by its verification to ensure adherence to the Collatz sequence properties.

The experiments were conducted on two different hardware configurations. All experiments in subsections~\ref{sec:analysis}, \ref{sec:verification}, \ref{sec:comparison}, and \ref{sec:scalability} were conducted on a MacBook Air equipped with an Apple M2 chip, 8GB of unified memory, and running macOS Sequoia Version 15.2. These experiments utilized Python 3.12.4 and Julia 1.11.2. The experiment in subsection~\ref{sec:execution_time} was conducted on Google Colab using an NVIDIA T4 GPU with Python 3.11.11.

\subsection{The Collatz Tree and Observed Patterns}

\begin{figure}[!htbp]
    \centering
    \includegraphics[width=\columnwidth]{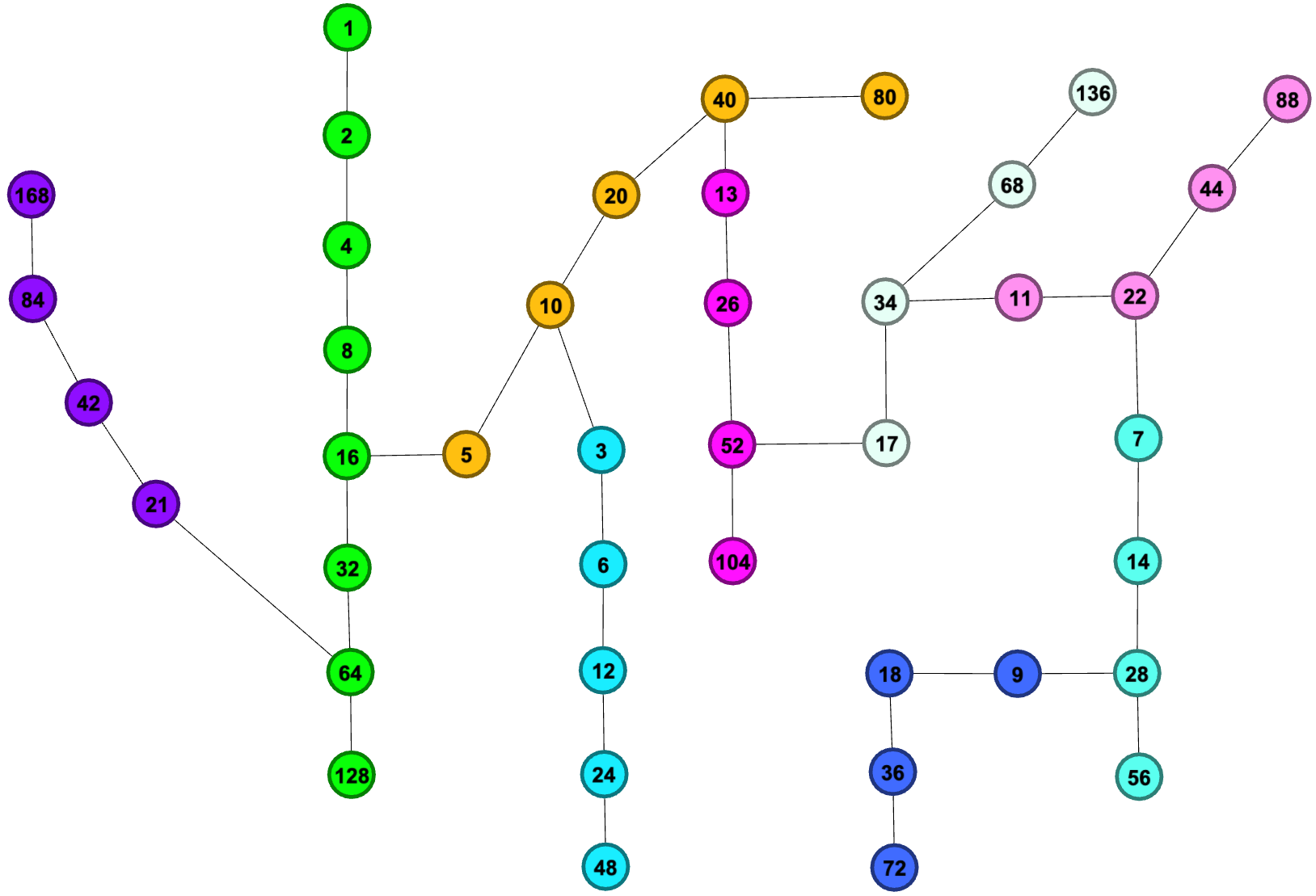}
    \caption{The Collatz tree with the base path (\(2^n\)) and sub-branches starting at \(6k+10\). Each sub-branch follows the pattern \(m \cdot 2^n\), where \(m\) is an odd number.}
    \label{fig:collatz_tree}
\end{figure}

The Collatz tree is a hierarchical representation of the inverse transformations of the Collatz sequence, starting from 1. As shown in Fig.~\protect\ref{fig:collatz_tree}, the structural patterns of the tree provide a foundation for developing efficient algorithms by identifying key paths and simplifying redundant computations.
   
\noindent \textbf{Base Branch:} The base branch starts at \(1\) and extends to infinity, following the pattern of powers of 2:
\[
\resizebox{\columnwidth}{!}{$
1 \to 2 \to 4 \to 8 \to 16 \to 32 \to 64 \to \ldots \to 2^n \ldots
$}
\]
\textbf{Sub-Branches:} All branches other than the base branch (sub-branches) also start with an odd number and extend to infinity as multiples of powers of 2. 
\[
\resizebox{\columnwidth}{!}{$
2m+1 \to 2(2m+1) \to \ldots \to (2m+1) \cdot 2^n \to \ldots
$}
\]
where \(m, n \geq 1\). \\

\noindent \textbf{Branch Points:} Branches occur throughout the Collatz tree at specific positions defined by \(6k + 10\) (\(k \geq 0\)). \\

\noindent \textbf{Special subtrees:} Odd multiples of 3 (\(3, 9, 15, 21, \ldots\)) form special subtrees which do not branch further. 
\[
\resizebox{\columnwidth}{!}{$
3(2m+1) \to 6(2m+1) \to \ldots \to 3(2m+1) \cdot 2^n \to \ldots
$}
\]
where \(m, n \geq 0\).

\subsection{Algorithm Development for Calculating Stopping Time}

The proposed algorithm leverages the structural properties of the Collatz tree and observed patterns in its branches to minimize redundant computations. \\

\noindent \textbf{Handling Odd Numbers:} If the current number is odd, it is the root of its branch. To move to the "previous branch" in the Collatz tree, compute the previous number using the equation:
\begin{equation}
\resizebox{0.8\columnwidth}{!}{$\text{previous Number} = 3 \times \text{current Number} + 1$}
\end{equation}
This computation gives the even number from which the current branch originated. \\

\noindent \textbf{Handling Even Numbers:} If the current number is even, it belongs to a branch rooted at an odd number. To find the root of the current branch, determine the largest power of 2 that divides the current number with a remainder of zero. This can be computed efficiently using the bitwise operation:
\begin{equation}
\resizebox{0.9\columnwidth}{!}{$\text{largest power of 2} = \text{current number} \land \text{-current number}$}
\end{equation}
where $\land$ is a bitwise AND operation. \\

\noindent Dividing the current even number by the largest power of 2 yields the root odd number:
\begin{equation}
\text{odd Root} = \frac{\text{current Number}}{\text{largest power of 2}}
\end{equation}

\noindent If the odd root equals 1, the base branch is reached, and the stopping time calculation is complete. \\

\noindent \textbf{Repeating the Process:} Alternate between handling odd and even numbers, reducing the current number until $n = 1$. This iterative process ensures efficient computation of the stopping time. \\

\noindent \textbf{Pseudocode:} The following pseudocode of the proposed algorithm outlines the steps to calculate the stopping time for a given number:

\begin{algorithm}[ht] 
\caption{CalculateStoppingTime}
\KwIn{$n$: Integer for which the stopping time is calculated}
\KwOut{$stopping\_time$: Total steps to reach 1}
\BlankLine
stopping\_time $\gets 0$\;
\While{$n \neq 1$}{
    \eIf{$n$ is odd}{
        $n \gets 3 \cdot n + 1$\;
        stopping\_time $\gets$ stopping\_time + 1\;
    }{
        largest\_power $\gets n \land -n$\;
        exponent $\gets \log_2($largest\_power$)$\;
        stopping\_time $\gets$ stopping\_time + exponent\;
        $n \gets n / $largest\_power\;
    }
}
\Return stopping\_time\ + 1;
\end{algorithm}

\noindent \textbf{Example Walkthrough for Algorithm:} We will compute the stopping time for \(n = 20{,}480\) using the given algorithm step-by-step: \\

\noindent \textbf{Initial Setup:} 
\begin{tabbing}
    \hspace{1em} \= \kill 
    \> Input: \(n = 20{,}480\) \\
    \> Initialize: \(\textnormal{stopping\_time} = 0\)
\end{tabbing}

\noindent\textbf{Iteration 1:} \(n = 20{,}480\) (even): 
\begin{tabbing}
    \hspace{1em} \= \kill 
    \> \(\textnormal{largest\_power} = 20{,}480 \land -20{,}480 = 4{,}096\) \\
    \> \(\textnormal{exponent} = \log_2(4{,}096) = 12\) \\
    \> \(\textnormal{stopping\_time} = 0 + 12 = 12\) \\
    \> \(n = \frac{20{,}480}{4{,}096}   = 5\)
\end{tabbing}

\noindent \textbf{Iteration 2:} \(n = 5\) (odd): 
\begin{tabbing}
    \hspace{1em} \= \kill
    \> \(n = 3 \times 5 + 1 = 16\) \\
    \> \(\textnormal{stopping\_time} = 12 + 1 = 13\)
\end{tabbing}

\noindent \textbf{Iteration 3:} \(n = 16\) (even): 
\begin{tabbing}
    \hspace{1em} \= \kill
    \> \(\textnormal{largest\_power} = 16 \land -16 = 16\) \\
    \> \(\textnormal{exponent} = \log_2(16) = 4\) \\
    \> \(\textnormal{stopping\_time} = 13 + 4 = 17\) \\
    \> \(n = \frac{16}{16} = 1\)
\end{tabbing}

\noindent \textbf{Termination:} \(n = 1\) 
\begin{tabbing}
    \hspace{1em} \= \kill
    \> \(\textnormal{stopping\_time} = 17 + 1 = 18\)
\end{tabbing}

\noindent Hence, the stopping time for \(n = 20{,}480\) is \(18\). While the state-of-the-art algorithms also compute this result in \(18\) iterations, the proposed algorithm achieves the same result using only \(3\) iterations by directly leveraging structural patterns in the Collatz tree. This optimization significantly reduces computational redundancy and highlights the algorithm's efficiency.

\subsection{Analysis of Proposed Algorithm}\label{sec:analysis}

\noindent To analyze the total iterations performed by the algorithm, consider the number $n = 48$ as shown in Fig.~\ref{fig:collatz_tree}. The traversal involves two sub-branches (starting with \(3\) and \(5\)) and the base branch. \\

\noindent For each sub-branch:
\begin{itemize}
    \item One iteration reduces the number to the root odd number that starts the sub-branch.
    \item One iteration moves up the tree to the parent sub-branch. 
\end{itemize}

\noindent In this case:
\begin{itemize}
    \item One iteration is used to reduce \(48\) to \(3\)
    \item One iteration moves 3 to the parent sub-branch starting with \(5\) (\(3\times 3+1=10\))
    \item One iteration is used to reduce \(10\) to \(5\)
    \item One iteration moves 5 to the base branch (\(5\times 3+1=16\))
    \item One iteration is used to reduce \(16\) to \(1\)
\end{itemize}

\noindent which gives a total of 5 iterations. \\

\noindent In general for $k$ sub-branches:
\begin{itemize}
    \item $k$ iterations are required to reduce the numbers in each sub-branch to their root odd numbers.
    \item $k$ iterations are required to traverse up the tree to the parent sub-branches.
    \item One iteration is required at the base branch to reduce the number to $1$.
\end{itemize}

\noindent Therefore, the total number of iterations is:
\begin{equation}
\text{Total Iterations} = 2k + 1
\label{eq:total_iterations}
\end{equation}

\noindent where $k$ is the number of sub-branches. \\

\noindent \textbf{Key Observations:}
\begin{itemize}
    \item Equation~\ref{eq:total_iterations} encapsulates the traversal of both even and odd numbers in the Collatz sequence.
    \item The value of $k$ depends on the structure of the Collatz tree and the placement of $n$ within it.
    \item Numbers in the base branch ($k = 0$) represent the best case, while numbers requiring traversal of many sub-branches correspond to the worst case. \\
\end{itemize}

\noindent \textbf{Best-Case Time Complexity:} The best case occurs when the input number $n$ is a power of 2 ($n = 2^k$). Numbers on the base branch of the Collatz tree (powers of 2) require no branching and are reduced directly to $1$ in a single iteration.

\paragraph*{Steps:}
\begin{enumerate}
    \item Compute \(\log_2(n)\) to determine the number of trailing zeros, which is performed in \(O(1)\) using optimized hardware instructions.
    \item The largest power of 2 dividing $n$ is $n$ itself, and $n / n = 1$ terminates the loop.
\end{enumerate}

\noindent \textbf{Stopping Time:} The stopping time is:
\[
\text{stopping\_time} = \log_2(n) + 1,
\]
where \(\log_2(n)\) accounts for trailing zeros and \(+1\) for the final step.\\

\noindent \textbf{Confirmation with $2k+1$:} In the best case (\(k = 0\)), the total iterations is \(2k+1=2(0)+1=1\), and this confirms that any number on the base branch reduces to \(1\) in \(O(1)\), without requiring any traversal through sub-branches. \\

\noindent \textbf{Time Complexity:} Each operation (\(\log_2(n)\), bitwise AND, division) is performed in \(O(1)\), leading to an overall best-case complexity of \(\mathbf{O(1)}\). \\

\noindent \textbf{Worst-Case and Average-Case Scenarios:}
To derive the worst- and average-case complexities, we considered the structure of the Collatz tree. The worst-case scenario occurs if the sequence contains an equal number of even and odd numbers, maximizing the number of branches (\(k\)). However, based on observations, this scenario is impossible due to the following observed patterns: \\

\begin{itemize}
    \item Even numbers dominate odd numbers in every branch due to rapid reductions by powers of \(2\)
    \item Branching only occurs at specific numbers: $$6k + 10, k\geq 0$$ 
    \item Odd multiples of \(3\) do not start new branches
    \item The tree structure indicates that branching occurs infrequently, and each branch has a large number of even reductions before reaching the next sub-branch. \\
\end{itemize}

\noindent Consequently, $k$, the number of sub-branches, is significantly smaller than the total number of even numbers in the sequence. Due to the challenges in deriving the exact sequence lengths and the precise number of odd numbers (or sub-branches) from the tree patterns, a computational approach was adopted to analyze the algorithm's behavior. \\

\noindent \textbf{Computational Approach:} 
The behavior of iterations and stopping times was analyzed using the proposed algorithm. The steps undertaken were as follows:
\begin{enumerate}
    \item The proposed algorithm was used to compute the stopping times and the number of iterations (\(2k + 1\)) for a range of inputs.
    \item Stopping times and iterations were analyzed over exponentially increasing ranges, such as \(10{,}000\), \(20{,}000\), up to \(5{,}120{,}000\).
    \item For each range, the best, average, and worst iterations were recorded to observe scaling behavior.
\end{enumerate}

\noindent The analysis, summarized in Table~\ref{tab:iterations_improved_clean}, presents the results for iterations across exponentially increasing ranges of inputs. The table highlights the best, average, and worst iterations for each range, providing insights into the algorithm's scaling behavior.

\begin{table}[h!]
\centering
\resizebox{\columnwidth}{!}{ 
\begin{tabular}{lccc}
\toprule
\textbf{Number of Inputs} & \textbf{Best Iterations} & \textbf{Average Iterations} & \textbf{Worst Iterations} \\ 
\midrule
10,000                    & 1                       & 56.90                       & 192                       \\ 
20,000                    & 1                       & 61.36                       & 204                       \\ 
40,000                    & 1                       & 65.68                       & 238                       \\ 
80,000                    & 1                       & 70.27                       & 258                       \\ 
160,000                   & 1                       & 75.00                       & 282                       \\ 
320,000                   & 1                       & 79.78                       & 328                       \\ 
640,000                   & 1                       & 84.59                       & 378                       \\ 
1,280,000                 & 1                       & 89.37                       & 392                       \\ 
2,560,000                 & 1                       & 94.15                       & 416                       \\ 
5,120,000                 & 1                       & 98.93                       & 444                       \\ 
\bottomrule
\end{tabular}
}
\caption{Best, Average, and Worst Iterations for Various Input Sizes}
\label{tab:iterations_improved_clean}
\end{table}

\begin{figure}[h!]
    \centering
    \includegraphics[width=\columnwidth]{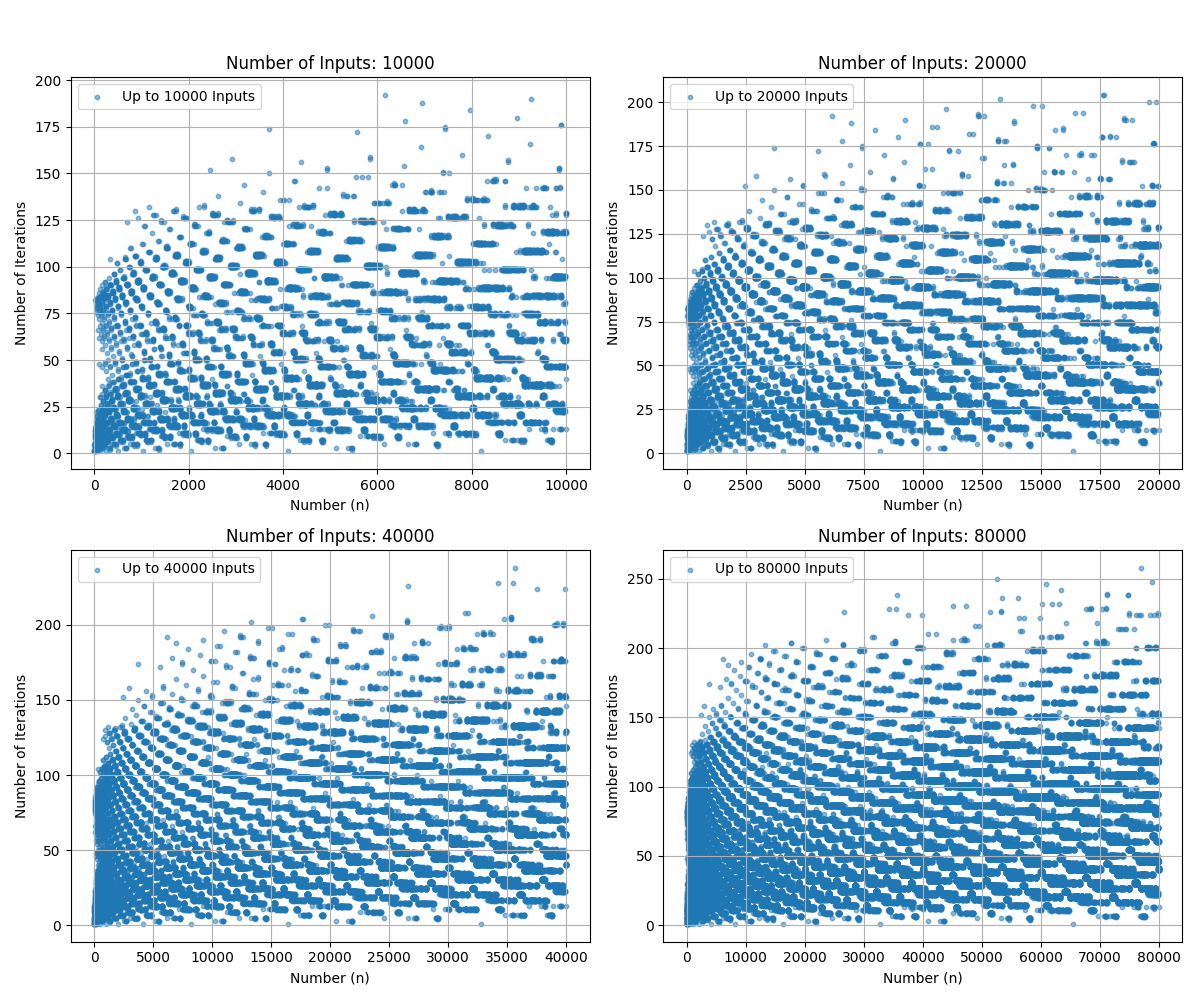}
    \caption{Scatter plots of the number of iterations versus numbers for selected input sizes. Each subplot represents a different input range, highlighting the trends in iterations as the input size increases.}
    \label{fig:iterations_grid}
\end{figure}

\noindent \textbf{Visualization:} Figure~\ref{fig:iterations_grid}, generated with Matplotlib, provides a visual representation of the iterations across different input sizes. The scatter plots demonstrate how iterations vary with input size, revealing the logarithmic growth of iterations with respect to the input numbers. Each subplot corresponds to a selected input range, offering a detailed view of the scaling behavior. \\  

\noindent \textbf{Average-Case Time Complexity:} To derive the average-case complexity, we analyzed the number of iterations for exponentially increasing input sizes. The observed average number of iterations for input sizes \(10{,}000\) to \(5{,}120{,}000\) indicated slow, logarithmic growth. \\

\noindent Specifically:

\begin{itemize}
    \item The average number of iterations grew by approximately \(5\) units as the input size doubled.
    \item This suggested a logarithmic relationship between the number of inputs and the average number of iterations.
\end{itemize}

\noindent Assuming the average number of iterations follows the form:
\[
f(n) = a \cdot \log_2(n) + b,
\]
we used two data points to calculate \(a\) and \(b\). \\

\noindent Given the data points:

\[
f(10,000) = 56.90, \quad \log_2(10,000) \approx 13.2877,
\]
\[
f(5,120,000) = 98.93, \quad \log_2(5,120,000) \approx 22.3181.
\]

\noindent we can express the relationships as a system of equations:

\[
\begin{aligned}
    56.90 &= a \cdot 13.2877 + b, \\
    98.93 &= a \cdot 22.3181 + b.
\end{aligned}
\]

\noindent Solving this system of linear equations yields:

\[
a \approx 4.65, \quad b \approx -4.91.
\]

\noindent Therefore, the average number of iterations as a function of \(n\) is given by:
\begin{equation}
f(n) = 4.65 \cdot \log_2(n) - 4.91
\label{eq:average_iterations}
\end{equation}

\noindent \textbf{Order of Growth}  
Equation~\ref{eq:average_iterations} confirms that the average number of iterations grows logarithmically with the input size. Consequently, the algorithm's average-case complexity is \(\mathbf{O(\log(n))}\), where \(n\) is the input size. This result aligns with the dominance of even reductions and the relatively shallow depth of sub-branches observed in the Collatz tree structure. \\

\noindent \textbf{Worst-Case Time Complexity:} The worst-case complexity was analyzed using the proposed algorithm, focusing on the maximum number of iterations (\(2k + 1\)) observed for exponentially increasing input sizes. \\

\noindent The results, summarized in Table~\ref{tab:iterations_improved_clean}, show that the worst-case number of iterations grow slowly with input size, consistent with logarithmic scaling. \\

\noindent For instance:

\begin{itemize}
    \item For \(n = 10{,}000\), the worst-case number of iterations were \(192\).
    \item For \(n = 5{,}120{,}000\), the worst-case number of iterations were \(444\).
\end{itemize}

\noindent To derive the worst-case complexity, we fit the observed data into the logarithmic form:
\[
f(n) = a \cdot \log_2(n) + b,
\]
where \(a\) and \(b\) are constants. \\

\noindent Using the data points:

\[
f(10{,}000) = 192, \quad \log_2(10{,}000) \approx 13.2877,
\]
\[
f(5{,}120{,}000) = 444, \quad \log_2(5{,}120{,}000) \approx 22.3181.
\]
\noindent We can express the relationships as a system of equations:

\[
\begin{aligned}
    192 &= a \cdot 13.2877 + b, \\
    444 &= a \cdot 22.3181 + b.
\end{aligned}
\]

\noindent Solving this system of linear equations yields:

\[
a \approx 28.00, \quad b \approx -180.17.
\]

\noindent The formula for the worst-case number of iterations as a function of \(n\) is therefore:
\begin{equation}
f(n) = 28 \cdot \log_2(n) - 180.17
\label{eq:worst_case_iterations}
\end{equation}

\noindent \textbf{Order of Growth:} 
Equation~\ref{eq:worst_case_iterations} confirms that the worst-case complexity scales logarithmically with the input size. Therefore, the algorithm's worst-case complexity is \(\mathbf{O(\log(n))}\), where \(n\) is the input size. \\

\noindent \textbf{Space Complexity:} The algorithm's space complexity is constant (\(\mathbf{O(1)}\)) across all cases due to its iterative nature. It uses a fixed number of scalar variables to store the current value of \(n\), the stopping time counter, and temporary intermediate results (e.g., largest power of 2). No additional data structures are required, and this remains true regardless of the input size or the number of iterations. 

\subsection{Verification of The Proposed Algorithm}\label{sec:verification}

A validation experiment was conducted to ensure the correct implementation of the proposed algorithm, comparing its results with those of the brute-force implementation. The validation process involved evaluating the stopping times of the first \(1\) billion numbers, confirming that the proposed algorithm accurately adheres to the Collatz sequence established in the theoretical analysis. \\

\noindent The experiment utilized OpenCL 1.2 to leverage parallel computing capabilities, thereby significantly reducing computation time. Parallelization was employed solely for accelerating the verification process and was not used in other experiments where the performance of the proposed algorithm was measured. Both the brute-force and proposed algorithms were executed in parallel for each number in the range \(1\) to \(1,000,000,000\). For each input, the stopping times computed by both algorithms were compared to ensure consistency and correctness across the entire range. \\

\noindent The results demonstrated a perfect match between the brute-force and proposed algorithms across all tested numbers. This successful validation confirms that the proposed algorithm preserves the fundamental properties of the Collatz sequence while optimizing computational efficiency. \\

\begin{figure}[ht]
    \centering
    \includegraphics[width=\columnwidth]{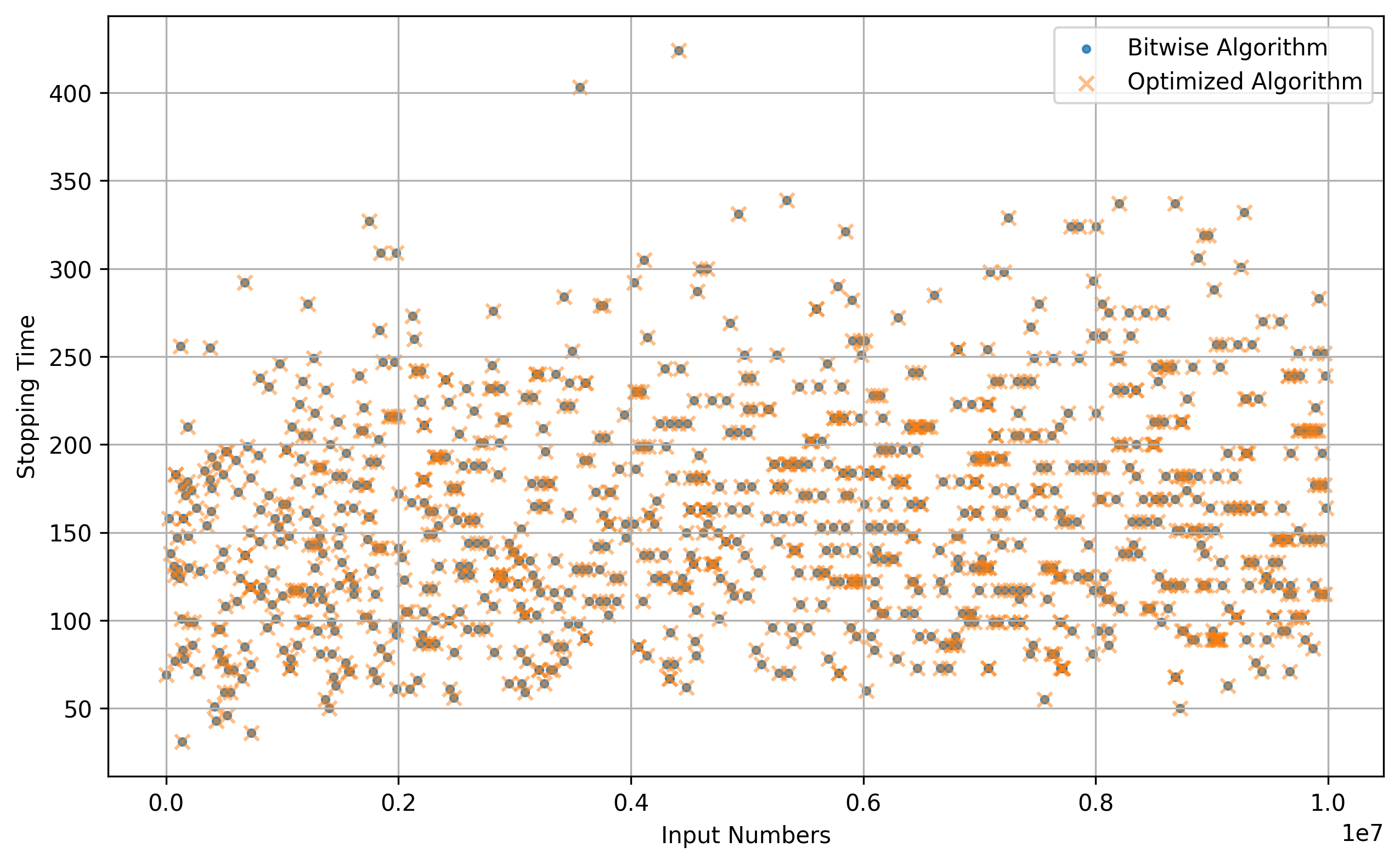}
    \caption{Comparison of stopping times for 1,000 randomly chosen numbers calculated by the brute-force and proposed algorithms.}
    \label{fig:verification_plot}
\end{figure}

\noindent To further illustrate the accuracy of the proposed algorithm, a subset of \(1,000\) randomly chosen numbers was plotted, comparing the stopping times calculated by both algorithms. As shown in Figure~\ref{fig:verification_plot}, the stopping times for the brute-force and proposed algorithms align perfectly, reinforcing the algorithm's correctness.

\subsection{State-of-the-Art Algorithms}

Verifying the Collatz conjecture for large numbers has seen limited progress despite several theoretical and computational efforts. Current computational approaches predominantly rely on brute-force methods or, in some instances, replace arithmetic operations with bit manipulations. This section focuses on analyzing two state-of-the-art algorithms that have advanced the verification of the conjecture. \\

\noindent \textbf{Algorithm 1:} The first algorithm, as presented in the paper \textit{Collatz conjecture for \(2^{100,000} - 1\) is true—Algorithms for verifying extremely large numbers}, extends the verification of the conjecture up to \(2^{100,000} - 1\) \cite{ren2018collatz}. This approach introduces logical computations and sequence encoding to optimize performance. The key features of this algorithm include:
\begin{itemize}
    \item Utilization of efficient encoding of the sequence, referred to as "code words" going forward, to reduce computational overhead.
    \item Replacement of traditional arithmetic operations with logical computations to improve speed.
    \item Scalability to handle extremely large numbers with reduced memory requirements. \\
\end{itemize}

\noindent The algorithm proposed in the paper replaces arithmetic computations in the \(3x+1\) operation with logical and bitwise operations. Additionally, it encodes the Collatz sequence into a compressed format we call "code words." These code words are designed to reduce the space required to represent the sequence by grouping each odd number and its immediate computations (\(3x+1\) and subsequent divisions) into a single symbol. The algorithm tracks two key metrics: \(U\), the number of \(3x+1\) computations, and \(D - U\), the additional number of divisions (\(x/2\)) not directly paired with \(3x+1\) \cite{ren2018collatz}. \\

\noindent \textbf{Examples from the Paper:}
For \(x = 7\), the Collatz sequence is:
\[
\begin{array}{l}
    7 \to 22 \to 11 \to 34 \to 17 \to 52 \to 26 \to 13 \to 40 \\
    \to 20 \to 10 \to 5 \to 16 \to 8 \to 4 \to 2 \to 1.
\end{array}
\]

\noindent The total sequence length is \(17\), and the algorithm tracks five \(3x+1\) computations (\(U = 5\)) and six additional division operations (\(D - U = 6\)). The sequence is encoded into the code word \texttt{"- - - 0 - 00 - 000"}, resulting in a code word length of \(11\), which matches the formula:

\begin{equation}
L_\text{code} = L_\text{seq} - N_\text{odd}, \label{eq:code_length}
\end{equation}

\noindent where:
\begin{itemize}
    \item \(L_\text{code}\) represents the length of the encoded sequence.
    \item \(L_\text{seq}\) denotes the length/stopping time of the sequence.
    \item \(N_\text{odd}\) indicates the total Number of Odd Numbers in the sequence.
\end{itemize}

\noindent For \(x = 63\), the Collatz sequence:
\[
63 \to 190 \to 95 \to 286 \to 143 \to \dots \to 1,
\]
has a total sequence length of \(108\). The algorithm tracks forty \(3x+1\) computations (\(U = 40\)) and twenty-eight additional division operations (\(D - U = 28\)). The sequence is encoded into the code word as follows resulting in a code word length of \(68\).

\[
\texttt{
\parbox{\columnwidth}{
"- - - - - - 000 - - 0 - - - 0 - - - - 0 - 00 - - - 0 - - 0 - - - - - - 00 - - - - 000 - 0 - 0 - 000 - 00 - - - 0000 - 000"
}
}
\]

\noindent \textbf{Shortcomings of the Approach:} A critical shortcoming of the algorithm is that the total number of computations (\(U + D\)) is equal to the sequence length. For \(x = 7\), the sequence length is \(17\), and the total computations \(U + D = 5 + 12 = 17\). Similarly, for \(x = 63\), the sequence length is \(108\), and the total computations \(U + D = 40 + 68 = 108\). This shows that the time complexity of the algorithm remains proportional to the sequence length:
\[
O(L_\text{seq})
\]

\noindent where \(L_\text{seq}\) is the length/stopping time of the sequence.\\

\noindent An additional limitation arises from the need to store the generated code words in secondary memory for verification purposes. This significantly increases the space complexity compared to simply storing the sequence length of a given number. For instance, storing the sequence length of \(7\) as a single number (\(17\)) is far more efficient than storing \(11\) symbols in the form of a code word. The code word length given in Equation~\ref{eq:code_length} shows that the space complexity of the algorithm is:

\[
O(L_\text{seq} - N_\text{odd})
\]
\noindent where \(L_\text{seq}\) is the length/stopping time of the sequence, and \(N_\text{odd}\) is the total Number of Odd Numbers in the sequence.\\

\noindent In summary, while the algorithm provides a novel perspective through logical operations and sequence encoding, the lack of improvement in time complexity and the higher space requirements for code words make it less optimal. \\

\noindent \textbf{Algorithm 2:} The second algorithm, presented in the paper \textit{Verification of Collatz Conjecture: An Algorithmic Approach}, modifies the brute-force approach by replacing arithmetic operations, such as division by 2 and \(3n + 1\), with bitwise operations ( \(n \gg 1\), \((n \ll 1) + n + 1\)) \cite{venkatesulu2020verification}. This optimization leverages hardware efficiency, processing each number iteratively while maintaining the same computational results. \\

\begin{algorithm}[ht] 
\caption{Collatz Length Computation Using Bitwise Operations \cite{venkatesulu2020verification}}
\KwIn{$n$: Integer for which the sequence length is computed}
\KwOut{$length$: Total length of the Collatz sequence for $n$}
\BlankLine
length $\gets 1$\;
\While{$n \neq 1$}{
    \eIf{$n \land 1 = 0$}{
        $n \gets n \gg 1$\;
    }{
        $n \gets (n \ll 1) + n + 1$\;
    }
    length $\gets$ length + 1\;
}
\Return length\;
\end{algorithm}

\noindent Algorithm~2, adapted from the methodology outlined in \cite{venkatesulu2020verification}, is presented here for further analysis. The algorithm maintains a time complexity of \(O(L_\text{seq})\), as each number in the sequence is processed exactly once, and achieves a constant space complexity of \(O(1)\) by utilizing only scalar variables to track the input and sequence length, thereby eliminating the need for additional data structures or sequence storage. \\

\noindent In contrast to the algorithm proposed by Ren et al.~\cite{ren2018collatz}, which stores the Collatz sequence in code words, Algorithm~2 computes the sequence length directly without storing intermediate results, achieving a constant space complexity of \(O(1)\), which is significantly more memory-efficient.

\section{Experiments}

\subsection{Comparison of Computational Efficiency Across Algorithms}\label{sec:comparison}

\noindent To compare the performance of the proposed algorithm against other state-of-the-art approaches, the results from Table 1 of the referenced paper \cite{ren2018collatz} are used as a baseline. These results include the input numbers, represented in decimal form as \(2^n - 1\) for varying \(n\), and the total number of computations, which are split into \(3x+1\) operations (denoted as \(U\)) and \(x/2\) operations (denoted as \(D\)). For consistency, the total computations (\(U + D\)) from the paper are included in the comparison. \\

\noindent In addition to these baseline values, the stopping times for each input number are recorded, and the total iterations for the proposed algorithm and the bitwise implementation of the brute-force algorithm (Algorithm~2, based on \cite{venkatesulu2020verification}) are calculated and included in Table~\ref{tab:comparison}. The table also shows the percentage improvement of the proposed algorithm over both \cite{ren2018collatz} and the bitwise implementation of the brute-force algorithm. The percentage improvement is calculated as follows:

\begin{equation}
\text{Improvement (\%)} = 
\left( 1 - 
\frac{\mathit{C_p}}
{\mathit{C_b}}
\right) \times 100
\label{eq:improvement}
\end{equation}

\noindent
Where:
\begin{itemize}
    \item $\mathit{C_p}$: Total iterations performed by the proposed algorithm.
    \item $\mathit{C_b}$: Total iterations performed by the benchmark algorithm.
\end{itemize}

\begin{table*}[ht]
\renewcommand{\arraystretch}{1.5} 
\centering
\caption{\centering Comparison of Total Iterations, Stopping Times, and Improvements for Different Algorithms}
\label{tab:comparison}
\begin{tabular}{|c|c|c|c|c|c|c|}
\hline
\textbf{Input} & \textbf{Stopping Time} & \multicolumn{3}{c|}{\textbf{Total Iterations}} & \multicolumn{2}{c|}{\textbf{Percentage Improvement}} \\ 
\cline{3-7}
               &                         & \textbf{Ren et al. \cite{ren2018collatz}} & \textbf{Bitwise} & \textbf{Proposed} & \textbf{Proposed over Ren et al.} & \textbf{Proposed over Bitwise} \\ \hline
\(2^{100} - 1\)    & 1465   & 1465   & 1465   & 1056   & 27.91\%   & 27.91\% \\ 
\(2^{500} - 1\)    & 6748   & 6748   & 6748   & 4834   & 28.35\%   & 28.35\% \\ 
\(2^{1000} - 1\)   & 12157  & 12157  & 12157  & 8632   & 29.00\%   & 29.00\% \\ 
\(2^{5000} - 1\)   & 67378  & 67378  & 67378  & 48262  & 28.37\%   & 28.37\% \\ 
\(2^{10000} - 1\)  & 134404 & 134404 & 134404 & 96252  & 28.36\%   & 28.36\% \\ 
\(2^{50000} - 1\)  & 667858 & 667858 & 667858 & 478040 & 28.40\%   & 28.40\% \\ 
\(2^{100000} - 1\) & 1344926 & 1344926 & 1344926 & 963206 & 28.36\%   & 28.36\% \\ 
\hline
\end{tabular}
\vspace{-1em} 
\end{table*}

\subsubsection*{Analysis of Results}

\noindent \textbf{Stopping Times:} The stopping times for all three algorithms are identical, confirming the correctness of the proposed algorithm in calculating the Collatz sequence. This equivalence highlights that the proposed optimizations do not alter the fundamental sequence traversal but instead reduce the computational overhead. \\

\noindent \textbf{Total Iterations:} The proposed algorithm consistently outperforms both algorithms terms of total iterations. For instance, for \(2^{100} - 1\), the proposed algorithm requires 1056 iterations while the other algorithms require 1465 iterations, demonstrating a significant reduction. \\

\noindent \textbf{Consistency Between Benchmark Algorithms:} The bitwise implementation of the brute-force algorithm (Algorithm~2, based on \cite{venkatesulu2020verification}) and the algorithm described in the paper \cite{ren2018collatz} produce identical total iterations for all input sizes. This consistency supports the earlier analysis, which established that both algorithms share the same time complexity. \\

\noindent \textbf{Percentage Improvement:} The proposed algorithm achieves a consistent improvement of approximately \(28\%\) across all input sizes compared to both the algorithm described in \cite{ren2018collatz} and the bitwise implementation of the brute-force algorithm (Algorithm~2, based on \cite{venkatesulu2020verification}). This consistency indicates that the optimizations introduced in the proposed algorithm scale effectively with larger inputs. \\

\noindent \textbf{Scalability:} As the input size increases exponentially (e.g., \(2^{100} - 1\) to \(2^{100000} - 1\)), the percentage improvement remains stable, indicating that the algorithm’s computational advantage is maintained regardless of the problem size.

\subsection{Execution Time Comparison of the Proposed and Bitwise Algorithms}\label{sec:execution_time}

\noindent To evaluate the computational efficiency of the proposed algorithm, a series of five experiments were conducted. These experiments compared the execution times of the proposed algorithm and the bitwise brute-force implementation(Algorithm~2), which served as a baseline for comparison. The algorithm described in \cite{ren2018collatz} was excluded from this analysis because, as demonstrated in previous results, it does not offer a significant advantage in time complexity compared to the bitwise approach. \\

\noindent The experiments were divided into two categories: general input tests (random numbers of varying sizes) and specific input tests (powers of two, multiples of three, and prime numbers). The general input tests assessed performance on arbitrary numbers, while the specific input tests aimed to highlight the strengths of the proposed algorithm when handling inputs with distinct patterns.

\subsubsection{General Input Tests:}

\begin{figure}[htbp]
    \centering
    \includegraphics[width=\columnwidth]{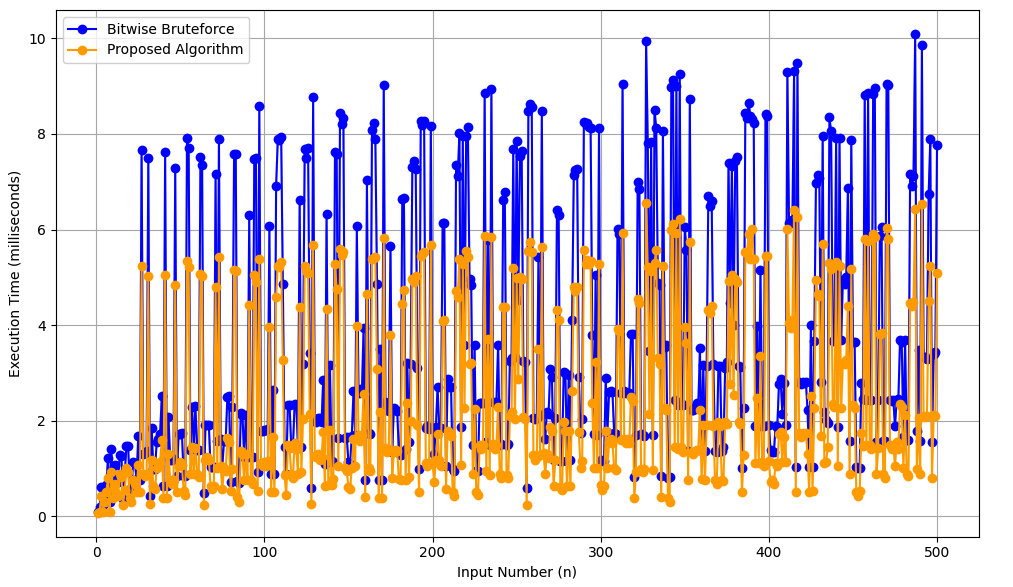}
    \caption{Execution time comparison for small random numbers.}
    \label{fig:small_random}
\end{figure}

The first experiment tested small random numbers (Figure~\ref{fig:small_random}), measuring performance on a small range of arbitrary inputs. The proposed algorithm consistently outperformed Algorithm~2 with lower execution times. \\

\begin{figure}[htbp]
    \centering
    \includegraphics[width=\columnwidth]{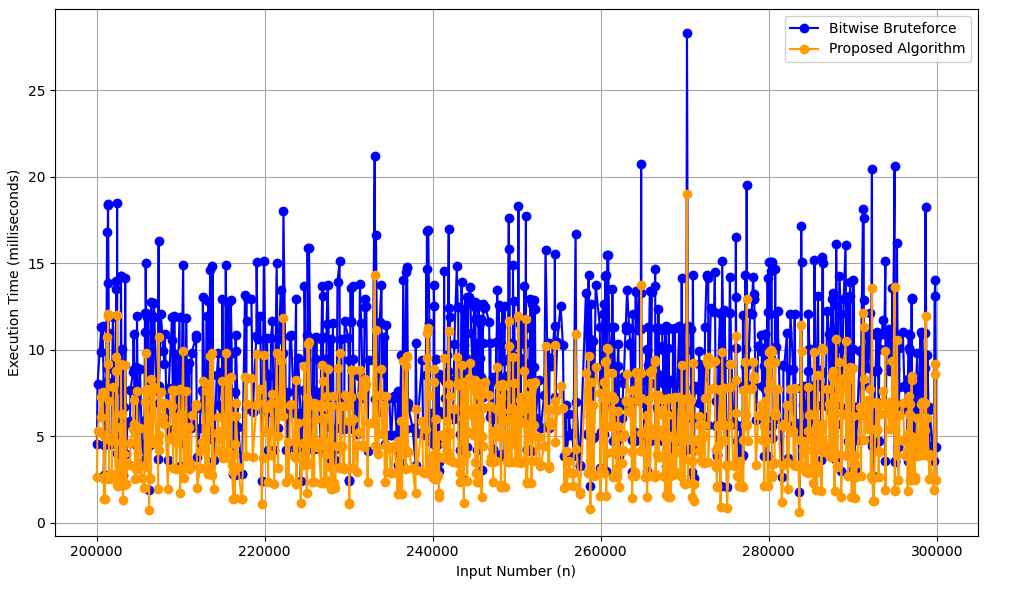}
    \caption{Execution time comparison for large random numbers.}
    \label{fig:large_random}
\end{figure}

\noindent The second experiment evaluated large random numbers (Figure~\ref{fig:large_random}) to assess scalability. Similar to the small inputs, the proposed algorithm maintained lower execution times, demonstrating efficiency across input sizes. \\

\subsubsection{Specific Input Tests:}

\begin{figure}[htbp]
    \centering
    \includegraphics[width=\columnwidth]{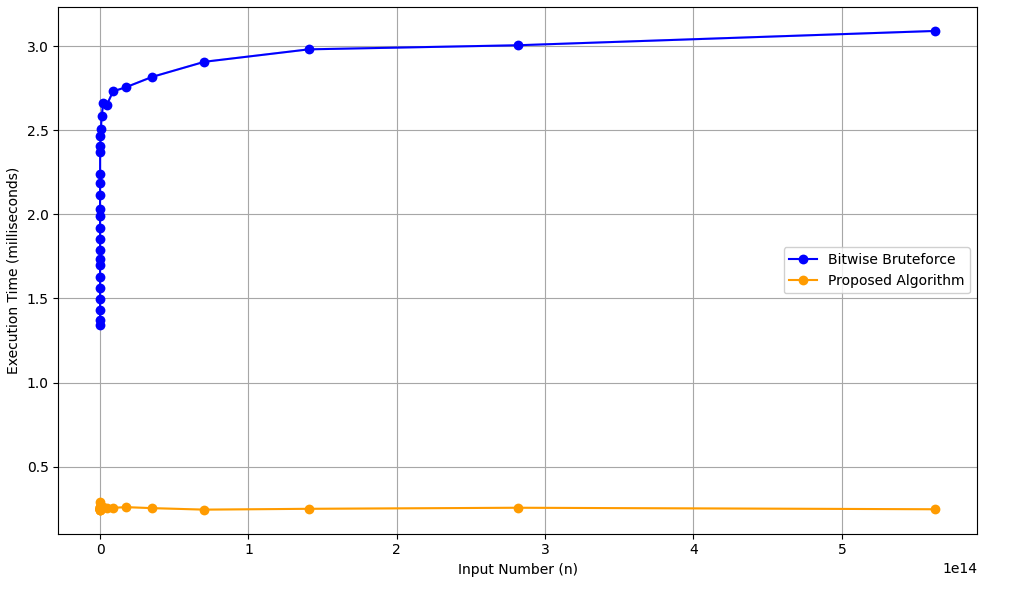}
    \caption{Execution time comparison for powers of two.}
    \label{fig:powers_of_two}
\end{figure}

The third experiment used powers of two (Figure~\ref{fig:powers_of_two}) to test performance on inputs with binary patterns. The bitwise brute-force algorithm(Algorithm~2) showed increased execution times due to trailing zeroes, while the proposed algorithm processed these inputs efficiently with minimal variance. \\

\begin{figure}[htbp]
    \centering
    \includegraphics[width=\columnwidth]{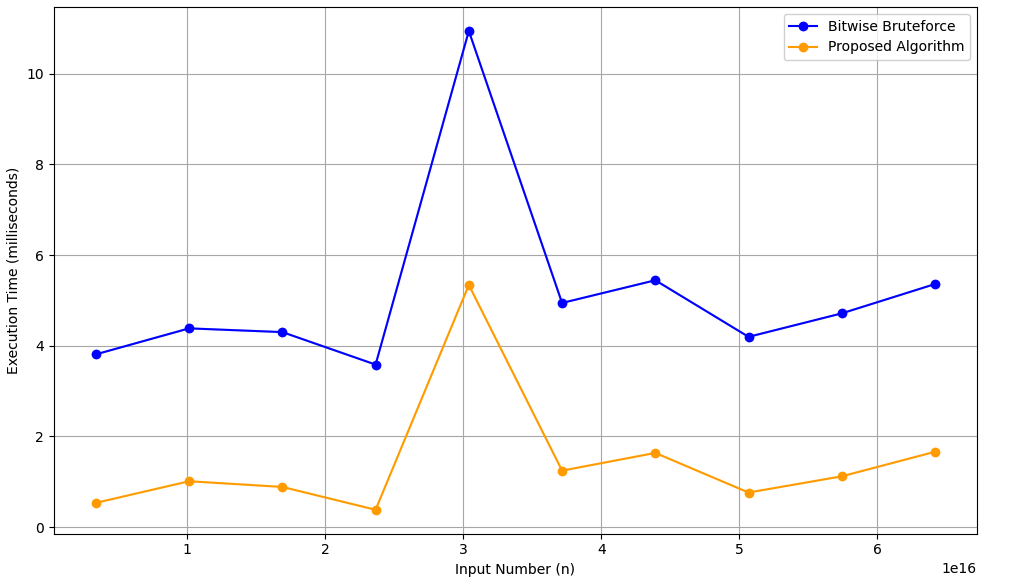}
    \caption{Execution time comparison for multiples of three.}
    \label{fig:multiples_of_three}
\end{figure}

\noindent The fourth experiment focused on multiples of three (Figure~\ref{fig:multiples_of_three}), selected because branches starting with these numbers terminate without forming sub-branches. This test evaluated the algorithms' handling of such simplified patterns, with the proposed algorithm showing significant execution time reductions compared to the bitwise implementation. \\

\begin{figure}[htbp]
    \centering
    \includegraphics[width=\columnwidth]{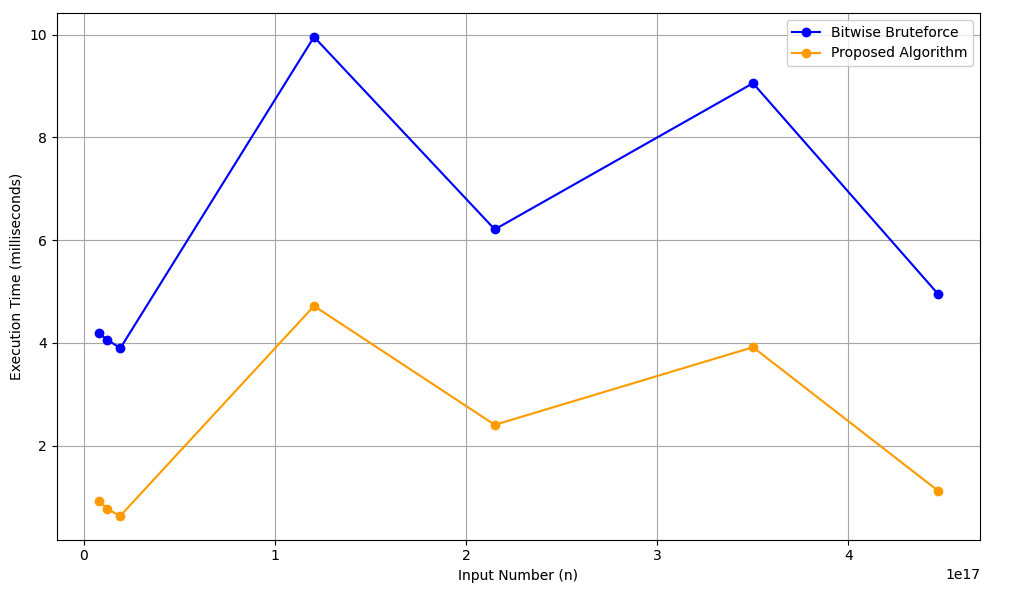}
    \caption{Execution time comparison for prime numbers.}
    \label{fig:prime_numbers}
\end{figure}

\noindent The fifth experiment analyzed prime numbers (Figure~\ref{fig:prime_numbers}), which often generate longer Collatz sequences. The bitwise algorithm exhibited high variability, while the proposed algorithm maintained consistently lower execution times, showcasing robustness. \\

\subsubsection{Summary of Results:}

The proposed algorithm consistently outperformed the bitwise briute-force implementation across all experiments, achieving lower execution times. Its optimizations in handling input patterns and reducing unnecessary operations highlight its efficiency for large-scale applications.

\section{Demonstrating Scalability: Verification Beyond \(2^{100,000}\)}\label{sec:scalability}

To evaluate the scalability of the proposed algorithm, an experiment was conducted to extend the verification of the Collatz conjecture beyond the previously verified upper bound of \(2^{100,000} - 1\) established by Ren et al. \cite{ren2018collatz}. The algorithm was tasked with computing the stopping times for numbers in the range:

\[
2^{100,000} \, \text{to} \, 2^{100,000} + 100,000,
\]

\noindent inclusive.

\subsection{Experimental Setup and Results}
The algorithm was implemented in Julia to leverage its support for arbitrary-precision arithmetic and faster execution compared to Python. Each number in the specified range was processed, and the stopping times were successfully computed. The results confirmed:
\begin{itemize}
    \item All numbers in the range have finite stopping times.
    \item No non-trivial cycles were discovered, consistent with the conjecture.
\end{itemize}

\subsection{Significance}
While the extended range of verification is relatively small compared to prior computational efforts, this experiment primarily serves to demonstrate the efficiency and scalability of the proposed algorithm. By successfully processing extremely large numbers without relying on additional techniques such as memoization, the results highlight the algorithm’s robustness and computational improvements. Large-scale verification of the conjecture is identified as a potential direction for future work, building on the foundations established by this research.

\section{Limitations, Scope, and Future Work}

The performance of the proposed algorithm is inherently constrained by the structure of the Collatz tree. As established in Equation~\ref{eq:total_iterations}, the total number of iterations depends on the number of branches, which are determined by the frequency of odd numbers in the sequence. Since all branches start with odd numbers, an increase in their occurrence leads to a higher branch count $k$, increasing the number of iterations. However, experimental results confirm a logarithmic relationship between input size and the number of iterations, even in the worst-case scenario, ensuring that the algorithm maintains scalability and efficiency across a wide range of inputs. 

The proposed algorithm is specifically designed for efficient stopping time computations based on the structural properties of the Collatz tree. Given that stopping time calculations are crucial for large-scale verification of the conjecture, this work primarily benefits applications requiring such computations. 

While the proposed algorithm efficiently computes stopping times, integrating memoization and parallelization can further enhance performance in large-scale verification processes. Parallelization allows the algorithm to run across multiple CPU or GPU cores, enabling concurrent computation of stopping times, while memoization reduces redundant calculations by storing previously computed values. These optimizations primarily benefit scenarios where multiple numbers need to be processed in bulk. 

Future research will focus on large-scale verification of the Collatz conjecture by employing memoization and parallelization to the proposed algorithm. By leveraging these techniques, significantly larger numeric ranges can be verified, further contributing to the computational exploration of the conjecture.

\section{Conclusion}

This research introduces an optimized algorithm for computing stopping times in the Collatz sequence, leveraging structural properties of the Collatz tree to minimize redundant operations. Experimental evaluations confirm that the proposed algorithm achieves a consistent 28\% reduction in computational iterations compared to state-of-the-art methods, demonstrating its efficiency across varying numeric ranges. Unlike prior approaches, the algorithm efficiently handles extremely large numbers without requiring auxiliary optimization techniques such as memoization or parallelization, reinforcing its scalability and robustness. 

A comprehensive experimental analysis was conducted to evaluate the algorithm's efficiency under different input conditions. The results highlight its ability to process small and large random numbers, powers of two, multiples of three, and prime numbers with improved execution times. Furthermore, by analyzing the worst-case and average-case complexity, this study establishes that the total number of iterations follows a logarithmic relationship with input size, ensuring stable performance even for large-scale computations. 

Beyond algorithmic optimization, this work provides insights into the structural behavior of the Collatz tree, contributing to a deeper understanding of sequence evolution. While the primary focus remains on optimizing stopping time calculations, the approach has broader implications in computational mathematics, including applications in sequence analysis, cryptography, and large-scale numerical verification. 

In summary, this study advances the state of computational approaches to the Collatz conjecture, offering a highly efficient algorithm for stopping time calculations. The proposed optimizations provide a foundation for extending computational verification efforts in number theory while opening new possibilities for algorithmic applications in theoretical and applied mathematics.

\section*{Acknowledgment}

The authors would like to acknowledge the use of ChatGPT \cite{openai2024chatgpt}, developed by OpenAI, for editing and grammatical checks in the preparation of this manuscript. Its assistance improved the clarity and readability of the text.

\bibliographystyle{IEEEtran}
\bibliography{references}

\EOD

\end{document}